\documentclass{phb-proc4-auth}

\usepackage{graphicx}
\usepackage{amssymb}

\begin{document}
\begin{frontmatter}


\journal{SCES '04}


\title{Magnetic-Field and External-Pressure Control of\\
Ferroelectricity in Multiferroic Manganites}

%
%
%
%
%
%

\author{H. Kuwahara\corauthref{1}},
\author{K. Noda},
\author{J. Nagayama}, and 
\author{S. Nakamura}
\address{Department of Physics, Sophia University, Tokyo 102-8554, Japan}
%

%
%
%
%


%
%
%
%

\corauth[1]{Corresponding Author: Dept.\ of Phys., Sophia Univ., 7-1 Kioi-cho, Chiyoda-ku, Tokyo 102-8554, Japan.  Phone\&FAX: +81-3-3238-3430, Email: h-kuwaha@sophia.ac.jp}


\begin{abstract}

We have investigated dielectric properties in a series of crystals of $R$MnO$_{3}$ ($R$ is a rare earth ion) under magnetic fields and quasihydrostatic pressure.  We have found that ferroelectric phase appeared in GdMnO$_3$ crystal below 13K.
We have confirmed that a small spontaneous polarization exists along $a$ axis ($P_a$) in the orthorhombic $Pbnm$ setting and that $P_a$ can be reversed by the dc electric field.  The dielectric anomaly due to the ferroelectric transition accompanied thermal hysteresis and lattice striction.  The ferroelectric transition temperature decreased with quasihydrostatic pressure. These results indicate that the ferroelectric transition is improper and is of the first-order displacement-type one.  $P_a$ was easily collapsed by application of magnetic field of 0.4T parallel to the spin-canting direction ($H \| c$) while it was enhanced parallel to the easy axis ($H \| b$).

\end{abstract}

%
%

\begin{keyword}

antiferromagnetism, ferroelectricity, phase transition

\end{keyword}


\end{frontmatter}

%
%
%
%
%

Systems with strongly coupled magnetic and electronic degrees of freedom have been attracting renewed interest since the advent of extensive studies of colossal magnetoresistive (CMR) manganites.
In a system of localized charge, effects of the coupling are demonstrated through a magnetocapacitive or magnetodielectric response, which is observed in several materials including YMnO$_3$~\cite{Katsufuji1}, EuTiO$_3$~\cite{Katsufuji2}, BiMnO$_3$~\cite{Kimura3}, and TbMn$_2$O$_5$~\cite{SWCheong}\@.
In spite of the intensive research for such a multiferroic material, i.e., compounds having (anti)ferromagnetic, (anti)ferroelastic, and/or (anti)ferroelectric properties, few multiferroics have been reported so far. 
The system investigated here, $R$MnO$_3$ ($R$ is a rare earth ion), is a parent antiferromagnetic (AF) Mott insulator of CMR manganites, which has a rich magnetic phase diagram showing the so-called Devil's flower with incommensurate and commensurate phases between a simple layered ($A$-type) and $E$-type AF phases with changing $R$ ions~\cite{Kimura2}.  Recently, Kimura {\it et al}.\ have reported on the intriguing discovery of magnetic ferroelectricity in the $R$=Tb crystal: the ferroelectric transition is observed at the incommensurate-commensurate (lock-in) magnetic one and ferroelectric polarization can be controlled by magnetic field~\cite{Kimura1}.

For the purpose of exploring intercrossing correlation between magnetic and electronic states, we have investigated dielectric properties in the series of crystals of $R$MnO$_{3}$ ($R$=La,Pr,Nd,Sm,Eu,Gd,Tb,Dy,Ho,Er, and Yb). We have performed the systematic measurements of dielectric constant, electric polarization, magnetization, and lattice striction as a function of temperature in magnetic fields and external quasihydrostatic pressure.  
Samples were grown by floating zone method and were cut along the crystallographic principal axes.

Figure 1 shows dielectric constant (a), spontaneous polarization (b), magnetization (c), and lattice striction (d) as a function of temperature for $a$-axis direction (in the orthorhombic $Pbnm$ setting) in GdMnO$_3$ crystal in the zero magnetic field.
As indicated by inversed triangles in (c), there are two distinguished phase transitions in this temperature ranges. One is the layered ($A$-type) AF transition of Mn $3d$ spins ($T^{\rm Mn}_{\rm N}$=20K for warming), where the weak ferromagnetic moment was observed. Mn $3d$ spins are slightly canted toward the $c$ axis, which is due to the Dzyaloshinsky-Moriya interaction.  The other is the ordering transition of Gd $4f$ spins that is reported in Ref.~7\@. Abrupt decrease in the weak ferromagnetic moment originates from the AF coupling between the ordered Gd $4f$ spins and the Mn $3d$ ones ($T^{\rm Gd\mbox{-}Mn}_{\rm N}$=13K for warming). 
As shown in (b), we have found that ferroelectric spontaneous polarization exists only along $a$ axis ($P_a$) below $T^{\rm Gd\mbox{-}Mn}_{\rm N}$.  The ferroelectric transition does not correspond to the $A$-type AF transition ($T^{\rm Mn}_{\rm N}$) or the incommensurate AF transition (42K~\cite{Kimura2}, not shown) of Mn.
The dielectric anomaly due to the ferroelectric transition at $T^{\rm Gd\mbox{-}Mn}_{\rm N}$ accompanies thermal hysteresis and lattice striction, as well as at $A$-type AF transition ($T^{\rm Mn}_{\rm N}$)\@. A large enhancement of $\varepsilon_a$ around $T^{\rm Mn}_{\rm N}$ was also detected.  We have found that the temperature for dielectric anomaly due to the ferroelectric transition decreased with increasing quasihydrostatic pressure.   These results indicate that the ferroelectric transition is improper and is of the first-order displacement-type one.  
Further investigation is, however, required to reveal the microscopic origin of the ferroelectricity, i.e., polar atomic displacements in this material.  
 
In order to clarify the coupling between magnetization and electric polarization, we have studied the effect of magnetic field on dielectric constant and polarization. 
As a result, the robustness of ferroelectric polarization against magnetic fields was sensitive to the magnetic-field direction. $P_a$ was easily collapsed by application of magnetic field of 0.4T perpendicular to the AF layers ($H \| c$) while it was enhanced parallel to the layers ($H \| b$). Thus obtained phase diagram in the magnetic field and temperature plane agrees well with the previously reported magnetic phase diagram~\cite{Hemberger}. The phase transition from ferroelectric to paraelectric seems to be induced by the metamagnetic transition of Gd sublattice or equivalently the destruction of the AF coupling between the Gd $4f$ spins and the Mn $3d$ ones.  
A gigantic magnetocapacitance effect was also observed around $T^{\rm Mn}_{\rm N}$\@.
These results are in strong contrast to the case of TbMnO$_3$~\cite{Kimura1}, in which the ferroelectric transition at the lock-in temperature is of the second order and $P_c$ is switched to $P_a$ (flopped) by application of magnetic field.

The family of compounds investigated here, $R$MnO$_3$, show the strong intercrossing correlation between magnetism and electric polarization.  This unusual intercrossing correlation makes these materials promising candidates for magnetically-recorded ferroelectric memory or electrically-recorded magnetic one.


\begin{figure}[htbp]
\centering
\includegraphics[width=60mm,clip]{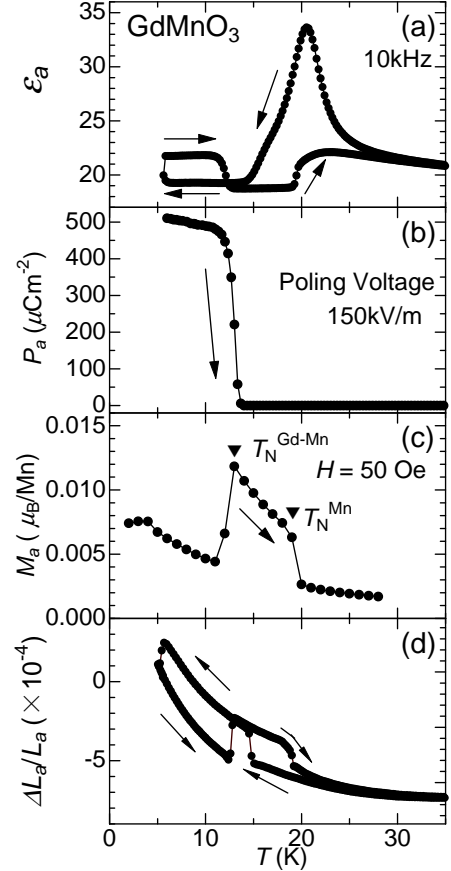}
\caption{Temperature dependence of (a) dielectric constant, (b) spontaneous polarization, (c) magnetization, and (d) lattice striction for $a$-axis direction in GdMnO$_3$ crystal in zero magnetic field.  The polarization was determined by integrating the measured pyroelectric current. After the sample was cooled in a poling voltage of 150kV/m, it was heated at a rate of 5K/min.  The lattice striction was measured by using a strain gauge.  Inversed triangles on the magnetization curve show $T_{\rm N}^{\rm Gd\mbox{-}Mn}$ or $T_{\rm N}^{\rm Mn}$ (see text).  Arrows represent a direction of thermal scans.}
\label{Fig1}
\end{figure}

%
%
%
%

%
%
%
%



\begin{thebibliography}{00}

\bibitem{Katsufuji1}
T. Katsufuji {\it et al.}, Phys. Rev. {\bf B 64}, 104419 (2001).

\bibitem{Katsufuji2}
T. Katsufuji and H. Takagi, Phys. Rev. {\bf B 64}, 054415 (2001).

\bibitem{Kimura3}
T. Kimura, S. Kawamoto, I. Yamada, M. Azuma, M. Takano, and Y. Tokura, Phys. Rev. {\bf B67} 180401(R) (2003).

\bibitem{SWCheong}
N. Hur, S. Park, P. A. Sharma, J. S. Ahn, S. Guha, and S-W. Cheong, Nature {\bf 429}, 392 (2004).

\bibitem{Kimura2}
T. Kimura, S. Ishihara, H. Shintani, T. Arima, K. T. Takahashi, K. Ishizaka, and Y. Tokura, Phys. Rev. {\bf B 68}, 060403(R) (2003).

\bibitem{Kimura1}
T. Kimura, T. Goto, H. Shintani, K. Ishizaka, T. Arima, and Y. Tokura, Nature {\bf 426}, 55 (2003).

\bibitem{Hemberger}
J. Hemberger, S. Lobina, H.-A. Krug von Nidda, N. Tristan, V.Yu. Ivanov, A.A. Mukhin, A.M. Balbashov, and A. Loidl, cond-mat/0311249.



\end{thebibliography}
\end{document}